\documentclass[aps,pra,letterpaper,twocolumn]{revtex4}

\usepackage{graphicx}
\usepackage{bm}
\usepackage{amssymb}
\usepackage{amsmath}
\usepackage{phoenician}

\usepackage{hyperref}

\newcommand{\skipthis}[1]{}

\newcommand{\dd}{\operatorname{d}\!}
\newcommand{\ee}{\text{e}}
\newcommand{\ii}{\text{i}}

\renewcommand{\eqref}[1]{Eq.~(\ref{#1})}
\newcommand{\figref}[1]{Fig.~\ref{#1}}
\newcommand{\secref}[1]{Sec.~\ref{#1}}

\newcommand{\eF}{\epsilon_{\rm F}}
\newcommand{\kF}{k_{\rm F}}

\begin{document}

\title{Ground state of a resonantly interacting Bose gas}

\author{J. M. Diederix}
\email{J.M.Diederix@uu.nl}
\author{T. C. F. van Heijst}
\author{H. T. C. Stoof}
\affiliation{
Institute for Theoretical Physics, Utrecht University,\\
Leuvenlaan 4, 3584 CE Utrecht, The Netherlands}

\date{\today}

\begin{abstract}
We show that a two-channel mean-field theory for a Bose gas near a
Feshbach resonance allows for an analytic computation of the
chemical potential, and therefore the universal constant $\beta$,
at unitarity. To improve on this mean-field theory, which
physically neglects condensate depletion, we study a variational
Jastrow ansatz for the ground-state wave function and use the
hypernetted-chain approximation to minimize the energy for
all positive values of the scattering length. We also show that
other important physical quantities such as Tan's contact and the
condensate fraction can be directly obtained from this approach.
\end{abstract}

\maketitle

\section{Introduction}\label{sec:intro}

In recent years, ultracold Fermi gases have been extensively studied near a Feshbach resonance, both theoretically and experimentally. Only more recently, the strongly interacting regime for an atomic Bose gas is beginning to be explored. The reason for this is that experiments are troubled by strong inelastic atom losses in this case. Nevertheless, a number of groups are now starting to carry out experiments in the strongly interacting regime near a Feshbach resonance \cite{papp2008bss,pollack2009eti,navon2011dtl}, and it is expected that significant results will soon be obtained. Theoretically, these systems are also challenging and some attempts toward an accurate description of the strong interaction effects that play a role here have already been made \cite{cowell2002cbg,song209gsp,lee2010usd}. In this paper we discuss another approach to study the ground state of resonantly interacting Bose gases.

As a first step and to discuss more transparently some of the important physics involved, we start this paper in \secref{sec:mf} with a mean-field description of an atomic Bose gas near a Feshbach resonance. This mean-field theory is based on a two-channel description containing both atoms and molecules, and has as a main approximation the neglect of depletion of the condensate. Using a two-channel model gives a finite energy for the Bose gas for all values of the scattering length $a$, also at unitarity, where the scattering length diverges. Moreover, near the Feshbach resonance the theory can be written in a universal form, which no longer depends on the specific details of the system. In this form it is even possible to find an analytic solution for the chemical potential at resonance.

However, this mean-field theory is not qualitatively reliable for large interaction strengths, since it neglects condensate depletion, which has significant effects on the energy. Therefore, we also study in \secref{sec:jastrowhnc} a variational Jastrow ground-state wave function combined with the hypernetted-chain approximation for the calculation of the ground-state energy. This approach has had great success in the strongly coupled helium liquids and we also show that, as desired, it reduces to the Bogoliubov theory in the weakly interacting limit. After a somewhat technical description in Secs.~\ref{sec:jastrow}--\ref{sec:energymin} on how to implement this approach, we show in Secs.~\ref{sec:larger}--\ref{sec:conddens} that it can be used to directly compute several important physical quantities. For instance, using this approach, the condensate fraction and the contact can be derived directly from the two-particle correlation function. As mentioned, the approach is variational. The total energy of the gas can be determined from the two-particle correlation function and the Jastrow factor, which are related to each other via the hypernetted-chain equation. The Jastrow factor, which determines the many-body ground-state wave function, is ultimately found by minimizing the energy. In \secref{sec:variational} we find that for the small and intermediate scattering length regime $n a^3<1$, where $n$ is the atomic density, this approach works very well, and also allows us to compute the contact and condensate fraction. However, the parametrization of the two-particle correlation function that we use here and that is inspired by the liquid helium literature, does not appear to work properly for larger scattering lengths and this
remains a topic for future work.

\section{Mean-field Theory}\label{sec:mf}

In ultracold dilute Bose gases, the interactions are usually completely determined by the $s$-wave scattering length $a$. However, the two-atom scattering problem can also contain bound states. In the case of a magnetic Feshbach resonance, the energy of these bound states depends on the externally applied magnetic field $B$. At certain values of this magnetic field, a new bound state can cross into the continuum of scattering states. At such a point there is a resonance in the scattering length, and the interaction appears infinitely strong in the $s$-wave channel.

In order to describe the many-body physics in such a system we start with an effective action for the atom field ($\phi_{\rm a}$) and molecule field ($\phi_{\rm m}$) that describes the bound state. This action can be derived from first principles \cite{duine2004amc}, and ultimately reads
\begin{align}\begin{split}
    \frac{S^{\rm eff}}{\hbar\beta V} = & -\mu\phi_{\rm a}^*\phi_{\rm a}^{\phantom{*}}+
        \phi_{\rm m}^*\left[\delta(B)-2\mu+\hbar\Sigma_{\rm m}\right]\phi_{\rm m}^{\phantom{*}}\\
        & +\frac{1}{2}T_{\rm bg}\phi_{\rm a}^*\phi_{\rm a}^*\phi_{\rm a}^{\phantom{*}}\phi_{\rm a}^{\phantom{*}}\\
        & +g\left[\phi_{\rm m}^*\phi_{\rm  a}^{\phantom{*}}\phi_{\rm a}^{\phantom{*}}+\phi_{\rm a}^*\phi_{\rm a}^*\phi_{\rm m}^{\phantom{*}}\right]\
        ,
\end{split}\end{align}
where $\mu$ is the chemical potential, $V$ is the volume, and $\beta = 1/k_{\rm B}T$ is the inverse temperature. For our purposes we can restrict ourselves to the zero-momentum and zero-frequency part. The atoms interact with each other in the so-called open channel with a strength $T_{\rm bg}$, proportional to the background scattering length  $a_{\rm bg}$ for the atoms, which is an experimentally known property of the specific Feshbach resonance of interest. The width of the resonance is determined by the atom-molecule coupling $g$ and is also known experimentally. The molecular energy depends on the external magnetic field through the self-energy $\hbar\Sigma_{\rm m}$ and via the magnetic detuning $\delta(B)\propto B-B_0$ from the resonance at the magnetic field $B_0$. For very broad resonances the interaction strength $g$ of the atoms with the molecules can, in principle, also depend on the magnetic field, but we neglect this feature here.

Minimizing the action gives rise to the following Gross-Pitaevskii equations for the atoms and molecules,
\begin{align}\begin{split}\label{eq:GP}
    \mu \phi_{\rm a}^{\phantom{*}} =\ & T_{\rm bg}|\phi_{\rm a}^{\phantom{*}}|^2\phi_{\rm a}^{\phantom{*}}+2g\phi_{\rm a}^*\phi_{\rm m}^{\phantom{*}}
 \\
    2\mu\phi_{\rm m}^{\phantom{*}} =\ & \left(\delta(B)+\hbar\Sigma_{\rm m}(2\mu - 2\hbar\Sigma_{\rm HF})\right)\phi_{\rm m}^{\phantom{*}}+g\phi_{\rm
    a}^2 ,
\end{split}\end{align}
where we have introduced the Hartree-Fock self-energy of the noncondensed atoms $\hbar\Sigma_{\rm HF}$. Its precise form is given below. The introduction of this self-energy is very important. Without the shift of the self-energy of the noncondensed atoms, the molecular condensate would always be unstable. Incorporating the Hartree-Fock self-energy makes sure that a condensate of molecules does not decay away immediately. In other words, the Hartree-Fock contribution to the self-energy makes sure that there exists a (metastable) equilibrium solution of the mean-field equations. Note that by elimination of the molecular field and considering the two-body limit, it is easy to show that the effective $T$ matrix of the atoms obeys the standard relation for the scattering length; that is, the effective scattering length $a(B)$ is related to the magnetic field via
\begin{align}\label{eq:gvsa}
    T_{\rm bg}-\frac{2g^2}{\delta(B)} \equiv \frac{4\pi a(B)\hbar^2}{m}\,.
\end{align}

For the broad Feshbach resonances of interest to us here, the molecular field and therefore the molecular density turn out to be very small, and we are allowed to put the atom density $n_{\rm a}$ equal to the total density $n$. As a consequence, the two-channel model now reduces to a single-channel model. The mean-field theory now reduces to solving the following three coupled equations
\begin{align}\label{eq:bogmol}\nonumber
    \mu &= n T_{\rm bg} + \frac{2n g^2}{2\mu - \delta(B) - \hbar\Sigma_{\rm m}(2\mu-2\hbar\Sigma^{\rm HF})} \\
    \hbar\Sigma_{\rm m}(E) &= -\frac{g^2m^{3/2}}{2\pi\hbar^3}\frac{\sqrt{-E}}{1+|a_{\rm bg}|\sqrt{-mE/\hbar^2}}\\ \nonumber
    \hbar\Sigma^{\rm HF} &=  2nT_{\rm bg} + \frac{4ng^2}{\hbar\Sigma^{\rm HF}+\mu-\delta(B) -\hbar\Sigma_{\rm m}(\mu-\hbar\Sigma^{\rm
    HF})},
\end{align}
where $a_{\rm bg}$ is the background scattering length associated with $T_{\rm bg}$. The first equation follows from the Gross-Pitaevskii equations in \eqref{eq:GP}, the second equation is the standard form of the molecule self-energy first derived in this context in Ref. \cite{duine2003jfr} and the third equation is the appropriate Hartree-Fock self-energy.

In this paper we are especially interested in the unitarity limit, which is the limit $a(B)\rightarrow\infty$. The physical properties of the atomic Bose gas are in this limit universal, which means that these properties do not depend on the specific details of the system, such as $a_{\rm bg}$ and $g$. This can be seen explicitly from the equation above. In the limit that $a\rightarrow\infty$ the background scattering length $a_{\rm bg}$ is irrelevant. Thus, we are allowed to take the limit $a_{\rm bg}\rightarrow 0$, while keeping $g^2/\delta(B)$ constant and still obeying \eqref{eq:gvsa}. Furthermore, the experimentally interesting case is a broad Feshbach resonance; we therefore take the limit $g\rightarrow\infty$ and $\delta(B)\rightarrow\infty$, while keeping the scattering length $a$ constant. In order to proceed further we introduce the Fermi momentum $\kF$ and Fermi energy $\eF$ instead of the density $n=\kF^3/6\pi^2$ and the mass $m/\hbar^2=\kF^2/2\eF$. We then end up with
\begin{align}
\begin{split}\label{eq:bogmoluni}
    \mu = & \frac{\eF}{3\pi}\frac{4\kF a}{1+\kF a\sqrt{-(\mu-\hbar\Sigma^{\rm HF})/\eF}},\\
    \hbar\Sigma^{\rm HF} = & \frac{\eF}{3\pi}\frac{8 \kF a}{1+\kF a\sqrt{-(\mu-\hbar\Sigma^{\rm
    HF})/2\eF}}, \end{split}
\intertext{where we used the following universal relations for
$a(B)$ and $\hbar\Sigma_{\rm m}(E)$:}\nonumber
    \frac{2g^2}{\delta(B)} = & -8\pi(\kF a)\frac{\eF}{\kF^3},\\ \nonumber
    \frac{\hbar\Sigma_{\rm m}(E)}{\delta(B)} =  & \kF
    a\sqrt{-\frac{E}{2\eF}}.
\end{align}
The former two equations give the chemical potential and the Hartree-Fock self-energy in units of the Fermi energy.

\begin{figure}
    \includegraphics[width=.9\columnwidth]{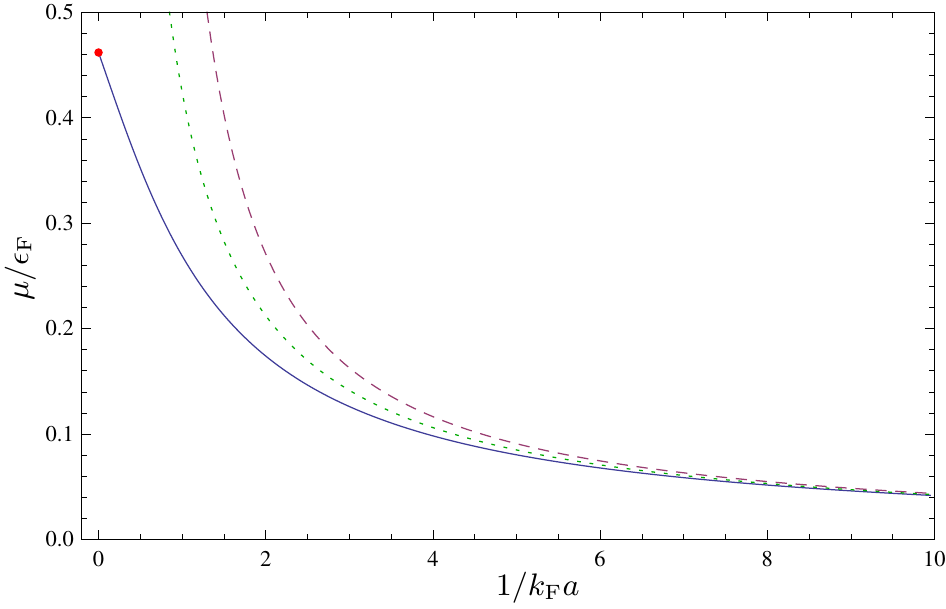}
    \caption{(Color online) The chemical potential $\mu$ in units of the Fermi energy $\eF$ as a function of the inverse scattering length $1/\kF a$. The solid line shows the mean-field result from \eqref{eq:bogmoluni}. The dashed line shows the Bogoliubov result from \eqref{eq:LDE} with LHY correction, while the dotted line is without this correction. Our two-channel mean-field approach stays finite in the unitarity limit, while the Bogoliubov theory diverges in that limit.}\label{fig:mfenergy}
\end{figure}

The two equations in \eqref{eq:bogmoluni} can be solved (in practice numerically) for any positive value of $a$. The result is shown in \figref{fig:mfenergy}. For small $a$, the relation for the chemical potential simply reduces to $\mu = 4\eF\kF a/3\pi = n T(a)$, which is the well known Gross-Pitaevskii expression for the small $a$ regime. As expected the Hartree-Fock self-energy then reduces to $\hbar\Sigma^{\rm HF}=2nT(a)=2\mu$. We can also solve the chemical potential explicitly in the unitarity limit. We then have
\begin{align}\begin{split}
    \frac{\mu}{\eF} = & \frac{4}{3\pi}\frac{1}{\sqrt{(\hbar\Sigma^{\rm HF} - \mu)/\eF}},\\
    \frac{\hbar\Sigma^{\rm HF}}{\eF} = & \frac{8}{3\pi}\frac{1}{\sqrt{(\hbar\Sigma^{\rm HF} - \mu)/2\eF}}\;.
\end{split}
\end{align}
Here we notice immediately that $\hbar\Sigma^{\rm HF} = 2\sqrt{2}\mu$, from which we can then easily solve for $\mu$ to obtain
\begin{align}\label{eq:muef}
    \mu = \sqrt[3]{\frac{4^2}{(3\pi)^2}\frac{1}{2\sqrt{2}-1}}\eF \simeq 0.4618 \eF\;.
\end{align}
For fermions, there is a similar relation, which is usually written as $\mu = (1+\beta)\eF$. In the specific case of fermions, the universal constant $\beta$ contains all the interaction effects and was found to be $\beta\simeq-0.58$ \cite{astrakharchik2004esf,carlson2005atf,partridge2006pps,zwierlein2006fsi}. It is customary to define a similar $\beta$ for bosons; the above mean-field theory gives $\beta\simeq-0.54$. This is just below an experimental lower bound set at $\beta > -0.56$ \cite{navon2011dtl}. Other theoretical analyses give varying results, namely, $\beta\simeq-0.34$ \cite{lee2010usd} and the upper bounds $\beta < 1.93$ \cite{cowell2002cbg} and $\beta < -0.20$ \cite{song209gsp}. It is remarkable that the fermionic value of $\beta$ is within these bounds: thus, it is not excluded that there is for this quantity no difference between fermions and bosons at unitarity. This might be anticipated in a one-dimensional situation; however, for a three-dimensional gas as considered here, this would be an interesting result indeed.

Another well-known mean-field result for the Bose gas energy is obtained from  Bogoliubov theory, which is an expansion in terms of the diluteness parameter $na^3$. This was already derived in the late 1950s in Ref.~\cite{lee1957} and reads
\begin{align}\label{eq:LDE}
    e = \frac{4\pi\hbar^2}{m a^2} na^3\left(1+\frac{128}{15\sqrt{\pi}}\sqrt{na^3}+\ldots\right)\;,
\end{align}
where $e$ is the energy per particle. The first term is the Gross-Pitaevskii (GP) result and the second term is known as the Lee-Huang-Yang (LHY) correction, and is due to condensate depletion resulting from quantum fluctuations.

The energy in Bogoliubov theory diverges in the limit $a\rightarrow\infty$; this is shown in \figref{fig:mfenergy} as the dashed line, which includes the LHY correction. The dotted line is the GP result without this correction. However, the energy for the mean-field theory of \eqref{eq:bogmoluni} stays finite, shown in \figref{fig:mfenergy} as the solid line. Of course, the energy of the unitary Bose system must be finite; thus the mean-field result of \eqref{eq:bogmoluni} describes this behavior correctly. However, from \figref{fig:mfenergy} can be concluded that quantum fluctuations, described by the LHY correction, are not properly incorporated in this theory. Thus, although qualitatively correct, the mean-field approach described in this section is probably not very reliable quantitatively. To improve on this we propose a different approach, based on a Jastrow wave function and the hypernetted-chain approximation which is discussed extensively in the rest of this paper.

\section{Jastrow and hypernetted-chain approximation}\label{sec:jastrowhnc}

In the previous section we have shown that with a simple mean-field theory, it is possible to capture the qualitative behavior of a Bose gas, where the energy stays finite when the scattering length diverges. However, this approach is probably not able to predict the energy reliably, since it excludes quantum fluctuations. We therefore propose an alternative approach in which we make a Jastrow ansatz for the wave function and use the hypernetted-chain approximation to compute correlations. This method was applied with great success in the field of strongly interacting helium \cite{smith1979gsb,feenberg1969}.

Since the Jastrow ansatz in combination with the hypernetted-chain approximation has been used successfully for some time now, there exists  a large amount of literature on the subject. However, in the field of ultracold atom gases, it is not used very often. We believe that these methods can be important for this field and we therefore briefly summarize the important relations and derivations in the sections below.

\subsection{Jastrow ansatz}\label{sec:jastrow}

The many-particle wave function can be a very complicated function of all the particle positions, but in the Jastrow approximation it is argued that the dominant correlation features are captured by the pair function or Jastrow factor $f(\bm r_1-\bm r_2)\equiv f(r_{12})$. The wave function is then,
\begin{align}\label{eq:jastrow}
    \Psi(\bm r_1 \ldots \bm r_N) = \prod_{i>j=1}^N f(\bm r_i - \bm r_j) \ .
\end{align}
In a homogeneous system this Jastrow factor only depends on the relative positions. This function goes to one on a length scale larger than the interparticle distance.

An important function in this description is the two-particle correlation function. It is defined as follows:
\begin{align}\label{eq:defineg}
    g(r_{12}) = \frac{N(N-1)}{ n^2}\frac{\int\dd \bm R_{12}|\Psi(\bm r_1, \ldots, \bm r_N)|^2}{\int\dd \bm R |\Psi(\bm r_1, \ldots, \bm r_N)|^2} \ .
\end{align}
Here $n$ is the density, $\int\dd \bm R = \int\dd\bm r_1\ldots\dd\bm r_N$ denotes the integration over all spatial coordinates, while $\int\dd \bm R_{12} = \int\dd\bm r_3\ldots\dd\bm r_N$ is the integration over all spatial coordinates except $\bm r_1$ and $\bm r_2$.

The energy of a system with a Jastrow wave function can be written in terms of the functions $f$ and $g$. The potential energy in terms of the Jastrow wave function is
\begin{align}\nonumber
    \left\langle V \right\rangle = \frac{\int\dd\bm R\Psi^*(\bm r_1,\ldots,\bm r_N)\sum_{i<j}^NV(r_{ij}) \Psi(\bm r_1,\ldots,\bm r_N)}{ \int\dd\bm R|\Psi(\bm r_1,\ldots,\bm r_N)|^2} \ ,
\end{align}
where $V(r)$ is the interparticle potential, which only depends on the distance between the particles. Using the particle-exchange symmetry of the wave function we can write this in terms of $g(r)$ as
\begin{align}
    \left\langle V \right\rangle =  \frac{1}{2} n^2 \int\dd\bm r_1\dd\bm r_2\ V(r_{12})g(r_{12}) \ .
\end{align}
The kinetic energy can be written as
\begin{align}\nonumber
    \left\langle T \right\rangle = -\frac{\hbar^2}{2m}\frac{\int\dd\bm R\Psi^*(\bm r_1,\ldots,\bm r_N)\sum_{i}^N\nabla_i^2 \Psi(\bm r_1,\ldots,\bm r_N)}{ \int\dd\bm R|\Psi(\bm r_1,\ldots,\bm r_N)|^2} \ ,
\end{align}
which can, again using the symmetry properties of the wave function, be written in terms of $f(r)$ and $g(r)$,
\begin{align}
    \left\langle T \right\rangle =-\frac{ n^2}{2} \frac{\hbar^2}{2m}\int \dd \bm r_1\dd \bm r_2\ g(r_{12})\nabla_{\bm r}^2\log f(r_{12}) \ .
\end{align}
Since we describe a homogeneous system, we can perform one more spatial integral, which gives a volume factor $V$. The total energy is now,
\begin{align}\label{eq:energy}
    e = \frac{1}{2}  n \int\dd\bm r\ g(r)\left(V(r) - \frac{\hbar^2}{2m}\nabla_{\bm r}^2\log f(r)\right) \ ,
\end{align}
where $e$ is again the energy per particle.

In this Jastrow ansatz for the wave function, everything in the system is determined by the Jastrow factor $f(r)$. However, many quantities, like the energy, are directly related to the two-body correlation function $g(r)$, but unfortunately the relation between $g(r)$ and $f(r)$ is very complicated. This relation [\eqref{eq:defineg}] contains as many integrals as there are particles, which clearly is unsolvable analytically. There are many approximation schemes to solve it, but many depend on small interactions, or correlation lengths. However, for bosons near a Feshbach resonance, we need an approximation scheme where these are large. The hypernetted-chain approximation, which is a diagrammatic cluster expansion, has proven to work also very well in the strongly interacting regime \cite{feenberg1969}. After we have established the relation between $f$ and $g$, we can solve for $f$ (or $g$) by minimizing the energy in \eqref{eq:energy}.

\subsection{Hypernetted-chain approximation}\label{sec:hnc}

With the Jastrow wave function we have a direct relation between the two-particle correlation function $g(r)$ and the Jastrow factor $f(r)$, but this relation contains, in the thermodynamic limit, an infinite number of integrals. These integrals cannot be solved analytically, but using the hypernetted-chain (HNC) approximation we can systematically evaluate them. The precise details of HNC can be read elsewhere, for example, in Ref. \cite{polls2002mdq}, but in the following we give a short derivation for completeness' sake to understand better the physics involved and introduce some useful notation.

We start by defining the cluster function $h(r) = f(r)^2 - 1$, which goes to zero quickly for large $r$, since $f(r)$ goes then to one. The two-particle correlation function $g$ can then in a natural way be written as a cluster expansion in terms of $h$,
\begin{align}
    &g(r_{12}) \propto \\\nonumber
     & \frac{\int\dd \bm R_{12}\left[1+\sum_{i<j}^Nh(r_{ij}) + \sum_{i<j}^N\sum_{k<l}^Nh(r_{ij})h(r_{kl})+\ldots\right]}
     {\int\dd \bm R\left[1+\sum_{i<j}^Nh(r_{ij}) + \sum_{i<j}^N\sum_{k<l}^Nh(r_{ij})h(r_{kl})+\ldots\right]}\;,
\end{align}
where the normalization constant is irrelevant for the discussion and is left out and the relative coordinates are defined as $r_{ij} = r_i - r_j$. These integrals are now written as an infinite sum of clusters of $h$, which each are a product of any number of $h$'s. These in turn can have different levels of complexity in terms of the integration variables. For example, $\int\dd r_3 h(r_{13})h(r_{23})$ is more complicated than $\int\dd r_3 \dd r_4 h(r_{13})h(r_{24})$. The idea behind hypernetted chain is to sum over an infinite amount of clusters selected by their complexity. When all complexities are taken into account, we end up with the exact result. However, in this paper we stick to the simplest set of clusters or diagrams, called nodal diagrams. This is referred to as  HNC/0. Since this is still a sum of an infinite amount of diagrams, the convergence of the approximation does not depend on the density or interaction strength to be small.

The nodal diagrams are all clusters of $h$ where the integral over a series of $h$'s only connects one $h$ to the next. Here `connect' means that we have an integral like $\int \dd r_3h(r_{13})h(r_{32})$, where $r_3$ `connects' the two cluster functions. We can construct an infinite set of these with the following recursion relation:
\begin{align}\begin{split}\label{eq:nodaldiag1}
    \mathcal{N}^{(0,1)}(r_{ab}) &= n\int\dd\bm r_1 h(r_{a1}) \mathcal{N}^{(0,1)}(r_{1b}) \\
      & \quad\quad\quad  + n\int\dd\bm r_1 h(r_{a1})h(r_{1b})\;,
\end{split}\end{align}
where $\mathcal{N}^{(0,1)}$ denotes the set of these simple nodal diagrams.

This set can, in turn, be used to generate an infinite amount of composite diagrams, which is simply all possible products between all the elements of $\mathcal{N}^{(0,1)}$. This we can write as
\begin{align}\begin{split}
    &\frac{1}{2!}\mathcal{N}^{(0,1)}(r_{ab})^2+    \frac{1}{3!}\mathcal{N}^{(0,1)}(r_{ab})^3+\ldots = \\
   &\quad\quad\quad\quad\quad\quad \exp{\left[\mathcal{N}^{(0,1)}(r_{ab})\right]}-\mathcal{N}^{(0,1)}(r_{ab})-1\;,
\end{split}\end{align}
where the numerical factors exactly cancel any double counting. This set can be extended even further by adding $h(r_{ab})\exp(\mathcal{N}^{(0,1)}(r_{ab}))$, leading to
\begin{align}\begin{split}\label{eq:compositediag1}
    &\mathcal{X}^{(0,1)}(r_{ab}) = \\ &\quad\quad\quad f^2(r_{ab})\exp{\left[\mathcal{N}^{(0,1)}(r_{ab})\right]}-\mathcal{N}^{(0,1)}(r_{ab})-1\;,
\end{split}\end{align}
where $\mathcal{X}^{(0,1)}$ is a set of all composite diagrams we can make with the set $\mathcal{N}^{(0,1)}$.

A lot more diagrams can be constructed by defining a set $\mathcal{N}^{(0,2)}$ that obeys \eqref{eq:nodaldiag1} but with $h(r_{ab})$ replaced by $\mathcal{X}^{(0,1)}(r_{ab})$. We can proceed naturally, and define a $\mathcal{X}^{(0,2)}(r_{ab})$ that obeys \eqref{eq:compositediag1} where $\mathcal{N}^{(0,1)}$ is replaced by $\mathcal{N}^{(0,2)}$. We can continue doing this, and in the limit where this procedure is followed an infinite number of times, we arrive at the following recursion relations:
\begin{align}\begin{split}\label{eq:hnc0a}
\mathcal{N}^{(0)}(r_{ab}) & =  n\int\dd\bm r_1 \mathcal{X}^{(0)}(r_{a1}) \mathcal{N}^{(0)}(r_{1b})  \\
   &\quad\quad\quad\quad  + n\int\dd\bm r_1 \mathcal{X}^{(0)}(r_{a1})\mathcal{X}^{(0)}(r_{1b})\;,
\end{split}
\end{align}
and
\begin{align}
\begin{split}\label{eq:hnc0b}
    g(r_{ab}) = &\ 1+\mathcal{N}^{(0)}(r_{ab})+\mathcal{X}^{(0)}(r_{ab}) \\
    =&\ f^2(r_{ab})\exp{\left[\mathcal{N}^{(0)}(r_{ab})\right]}\;,
\end{split}
\end{align}
where $\lim_{k\rightarrow\infty}\mathcal{N}^{(0,k)}=\mathcal{N}^{(0)}$ and $\lim_{k\rightarrow\infty}\mathcal{X}^{(0,k)}=\mathcal{X}^{(0)}$. This latter equation relates the two-particle correlation function $g$ to the Jastrow factor $f$, which is what we needed. This selected set of diagrams used to compute $g$ is called HNC/0. In order to include more(all) contributing diagrams we would have to include also more(all) elementary diagrams in \eqref{eq:compositediag1}, in addition to the nodal diagrams. However, this HNC/0 approximation contains already a lot of important information, as was shown by the calculations on strongly interacting helium.

The relation between $f$ and $g$ in \eqref{eq:hnc0b} can be solved for $f$ as
\begin{align}\label{eq:fintermsofg}
    \log f^2(r) = \log{g(r)} - \mathcal{N}^{(0)}(r) \ .
\end{align}
The usefulness of this equation follows from the fact that the function $\mathcal{N}^{(0)}(r)$ can also be related to the two-particle distribution function. To do this we first define the structure factor $S(k)$,
\begin{align}\label{eq:defenitionofs}
    S(k) = 1 +  n \int\dd\bm r\ \ee^{\ii \bm k\cdot\bm r}\left(g(r)-1\right) \ .
\end{align}
Note that from the definition of $g$ in \eqref{eq:defineg} it follows that $S(0) = 0$. The integral relations in Eqs.~\ref{eq:hnc0a} and \ref{eq:hnc0b} can be written as algebraic equations after a Fourier transformation. These equations are then easily solved and we get $\mathcal{N}^{(0)}(k)$ in terms of $S(k)$,
\begin{align}
    \mathcal{N}^{(0)}(k) = \frac{\left(S(k)-1\right)^2}{S(k)} \ ,
\end{align}
where $\mathcal{N}^{(0)}(k)$ is the Fourier transform of $\mathcal{N}^{(0)}(r)$. In the HNC approximation, the Jastrow factor $f(r)$ is thus completely determined in terms of the two-particle correlation function $g(r)$.

In this paper, we will vary the function $g$ and then calculate the energy. For this we need $f$, which we can calculate using the above equations. Since we need some complicated shape for $g$, it is not possible to analytically perform the Fourier transformations. Calculating the energy thus involves a few steps. First, when we have a $g$, we calculate the structure factor $S(k)$ by numerically Fourier transforming this $g$. Second, we calculate $\mathcal{N}^{(0)}(k) $ and numerically inverse Fourier transform back. Third, we calculate $f$ with which we can compute the energy.

\subsection{Energy minimization}\label{sec:energymin}

With the relations that follow from the HNC approximation, we can write the energy of the system in terms of only the two-body correlation function. To get the ground-state wave function, we have to minimize this energy with respect to this function. We first write down an analytic expression for this minimization condition, but it turns out to be hard to solve this relation in practice. It is much more convenient to numerically minimize the energy.

We have a relation for the energy in terms of $f$ and $g$ in \eqref{eq:energy}, and in combination with \eqref{eq:fintermsofg} we can write this in terms of $g$ only. Taking the functional derivative of the energy with respect to $g(r)$, or more conveniently $\sqrt{g(r)}$, and putting that to zero gives the following differential equation for $g$
\begin{align}\label{eq:minimizeg}
    \left\{-\frac{\hbar^2}{2m}\nabla^2+\left[V(r
    )+\omega_0(r)\right]\right\}\sqrt{g(r)} = 0 \;,
\end{align}
where $\omega_0(r)$ is defined as the inverse Fourier transform of
\begin{align}\label{eq:defomega}
    \omega_0(k) = -\frac{\hbar^2k^2}{4m}\left(S(k)+1\right)\left(1-\frac{1}{S(k)}\right)^2\ .
\end{align}
The minimization equation \eqref{eq:minimizeg} has the form of a simple Schr\"odinger equation for $\sqrt{g}$ where $\omega_0(r)$ acts as an effective induced potential that takes the presence of the entire medium into account. This may seem as a simple-to-solve equation, but recall that $\omega_0(r)$ contains $g$ in a very non-linear way.

Solving this differential equation for $g$ numerically turns out to be very hard. Small numerical errors in $g$ trigger solutions of the differential equation that are not physical, that is, these solutions are not normalizable. By explicitly varying $g$ to minimize the energy, these problems can be circumvented.

\subsection{Asymptotic behavior}\label{sec:larger}

Even without minimizing the energy we can say something about the shape of the two-body correlation function $g$. Let us first study the case for small scattering length $\kF a$. In this regime, the known Bogoliubov dispersion relation can be related to the structure factor $S(k)$. This relation follows from the dispersion relation from Bijl-Feynman theory, which reads
\begin{align}\label{eq:bfdispersion}
    E(k) = \frac{\hbar^2 k^2}{2m S(k)} \ .
\end{align}
In Bogoliubov theory, the dispersion relation is given by,
\begin{align}\label{eq:dispersion}
    E(k) = \sqrt{\epsilon_{\bm k}^2 + 2 n T(a)\epsilon_{\bm k}}\ .
\end{align}
Here, $\epsilon_{\bm k} = \hbar^2 k^2/2m$ and $T(a) = 4\pi\hbar^2a/m$. When we combine \eqref{eq:bfdispersion} and \eqref{eq:dispersion} we get the following for $S$,
\begin{align}
    S(k) &=  \frac{k^2}{\sqrt{k^4 + 16\pi a n k^2}}\\\nonumber
         &= \frac{k}{\sqrt{16\pi a n}} - \frac{k^3}{2(16\pi a n)^{3/2}} + \mathcal{O} (k^5)\;.
\end{align}
Thus, for small $k$ we have $S(k)= \hbar k/2mc$ with $c=\sqrt{4\pi a n}$ the speed of sound of the medium.

As was pointed out before, the structure factor is related to the two-body correlation function $g$. Since we know the behavior of $S(k)$ for small $k$, we can deduce the large-$r$ behavior of $g$. Using the asymptotic Fourier transform we get for large $r$ that
\begin{align}\label{eq:gtail}
    g(r\rightarrow\infty) = 1 - \frac{\hbar^2}{2\pi^2nmc}\frac{1}{r^4} \ .
\end{align}
This result holds only for small $a$, however, for large $a$ one still expects to find a linear dispersion relation for small $k$. This means that the $1/r^4$ tail will have a different prefactor, but should still be there in the unitarity limit. The tail of the trial functions for $g$, which we use in the variational calculation, will therefore be of that form. From the prefactor we can determine the speed of sound.

The Jastrow factor $f$ is completely determined by $g$ (and $S$) and the large-$r$ tail of this function is thus also known
\begin{align}
    f(r\rightarrow\infty) = 1 - \frac{mc}{\pi^2n\hbar}\frac{1}{r^2} \ .
\end{align}
These limits also tell us something about the large-$r$ behavior of the effective induced potential in \eqref{eq:defomega},
\begin{align}
    \omega_0(r\rightarrow\infty) = -\frac{3\hbar^2}{4m\sqrt{an^3\pi^5}}\frac{1}{r^6}\ .
\end{align}
This result is consistent with the analytic minimization equation in \eqref{eq:minimizeg}, since (when we put $V$ to zero) the two limits for both $g$ and $\omega_0$ exactly solve this differential equation.

\subsection{Contact}\label{sec:contact}

The small-$r$ behavior of the two-body correlation function can be related to what is called the \emph{contact}, denoted by $\mathcal{C}$. The contact was recently derived to be a general feature in strongly interacting Fermi systems by Tan \cite{tan2008esc,tan2008lmp}, in a sequence of papers published in 2008. Since Tan's derivation is not based on the statistics of the particles, it was pointed out by Combescot \emph{et al.}~\cite{combescot2009pdt} that the relations also hold for Bose statistics and are hence applicable to Bose gases as well. The quantity $\mathcal{C}$ is part of a series of various exact and universal relations which therefore also hold for strongly correlated gases. When applied to Bose gases, Tan's main theorem, which he calls the ``adiabatic sweep theorem'', states
\begin{align}\label{eq:tancontact}
    -\frac{\dd\,(n\;e)}{\dd\,(1/a)} = \frac{\hbar^2}{m}\frac{\mathcal{C}}{8\pi},
\end{align}
here $e$ is the energy per particle of the gas. It is striking that this is such a simple, exact and universal relation. The contact $\mathcal{C}$ turns out to be independent of the short-range interactions, except for the scattering length $a$. In general, it is a constant which is expected to remain finite for all values of the scattering length. Let us consider the well-known low-density expansion, or Bogoliubov theory for the ground-state energy \eqref{eq:LDE}. When we apply Tan's theorem to this energy expression, we find an approximation for the contact of a Bose gas for small scattering length to first order,
\begin{align}\label{eq:ldecontact}
    \mathcal{C} = (4\pi na)^2\left(1+\frac{64}{3\sqrt{\pi}}\sqrt{na^3}\right)\,.
\end{align}
It was shown \cite{schakel2010trd} that this equation for $\mathcal{C}$ can be derived independently of \eqref{eq:tancontact}, from which can be seen that Tan's relations agree with Bogoliubov theory.

The relation for the contact in \eqref{eq:ldecontact} does not have a finite limit at a Feshbach resonance when $a$ goes to infinity, since Bogoliubov is only valid for small $na^3$. However, the method proposed in this paper does have a finite limit. It is in general possible to find $\mathcal{C}$ in terms of the two particle distribution function $g(r)$, which in the context of HNC/0 is of great use.

For small $r$, the behavior of the two-particle distribution function $g(r)$ is dominated by the interaction of only two particles, since in a dilute gas, the rest of the particles are far away. The function $\sqrt{g}$ is therefore proportional to the two-particle wave function $f_2(r)$, which is the solution of the two-particle Sch\"odinger equation (see \secref{sec:feshbach}). For small $r$, but outside the range of the interaction, this function behaves as $f_2(r)\simeq 1-\frac{a}{r}$. The two-body distribution function $g$ is thus for small $r$ proportional to $f_2^2$, with a proportionality factor we call $Z$. Thus, we have for small $r$,
\begin{align}
    g(r) \simeq Z |f_2(r)|^2 \simeq Z a^2 \left(\frac{1}{r^2} - \frac{2}{ar} \right)\;.
\end{align}
The proportionality constant $Z$ is related to the contact through $Z = \mathcal{C} / 16 \pi^2 n^2 a^2$. We thus have for the contact
\begin{align}\label{eq:contactfromg}
    \mathcal{C} \simeq 16\pi^2n^2a^2\frac{g(r)}{|f_2(r)|^2} \;,
\end{align}
which can be calculated directly from the HNC solutions. We calculate the energy as a function of the scattering length, and as a result, we are also able to use the original expression in \eqref{eq:tancontact} to compute $\mathcal{C}$, which we compare to the results from \eqref{eq:contactfromg}.

In the experiments that a number of groups a trying to perform, one of the biggest challenges is the severe losses of the particles in the trap. An important consequence of the wave function renormalization factor $Z$, which is related to the contact, is that it also affects the three-body collision terms, which is what determines the particle loss rate of the Bose gas in a trap. This particle-loss is governed  by the relation
\begin{align}\label{eq:loss}
    \frac{\dd n}{\dd t} = L\ n^3\ ,
\end{align}
where $L$ determines the loss rate, and the power of $n^3$ reflects that three-body collisions are needed to obey the conservation laws. Since the wave function amplitude changes with $Z<1$ at small distances, the loss rate $L$ is multiplied with $Z^3$ due to many-body effects. Since the contact, and hence $Z$, can change significantly near the Feshbach resonance, this will have great effect on the losses in experiments.

\subsection{Condensate density}\label{sec:conddens}

An important physical quantity is the condensate fraction, denoted by $n_0$. This is the density of particles which are in the zero-momentum state and form a Bose-Einstein condensate. Conversely, there is a density of particles which are not in the condensate, due to (quantum) \emph{depletion}. This density is typically nonzero even at zero temperature, an effect which is solely due to interactions. In this section, we follow the lines of Ristig \emph{et al.}, see Refs.~\cite{ristig1975cfm,ristig1976dmq,ristig1979lob}.

We first consider the one-body density matrix for the system of $N$ bosons, given by
\begin{align}
n(r_{11'})=N\frac{\int \dd \bm R_1 \Psi^{*}(\bm r_1,\ldots,\bm r_N)\Psi(\bm r_1',\ldots,\bm r_N)}{\int \dd \bm R_1 |\Psi(\mathbf{r}_1,\ldots,\mathbf{r}_N)|^2},
\end{align}
where $\Psi$ is the wave function for the system. This can be written in a convenient notation as
\begin{align}\label{eq:densmatrix}
    n(r) = n_0 \ee^{-Q(r)}\ .
\end{align}
The one-body density matrix has the well-known properties that $n(0) = n$ and $n(r\rightarrow\infty)=n_0$, which in terms of $Q$ means $Q(r\rightarrow\infty)=0$ and  $n_0 = n\ee^{Q(0)}$. Using a cluster expansion, similar to HNC/0, this $Q(r)$ can be computed. The details of this computation can be found in Refs.~\cite{ristig1975cfm,ristig1976dmq,ristig1979lob}, but in the following we give a brief outline of it.

The most insightful approach to the calculation of $Q(r)$ is the method proposed by Feenberg \cite{feenberg1974gsi}. The ground-state Jastrow wave function, written in \eqref{eq:jastrow} is not necessarily properly normalized. A trial wave function that can be properly normalized, and was proposed by Feenberg, is given by
\begin{align}\label{eq:feenberjastrow}
    \Psi^{(N)}(\mathbf{r}_1,\ldots,\mathbf{r}_N)=\ee^{-\frac{\lambda N}{2}}\left(\frac{n}{N}\right)^{N/2}\prod^{N}_{i<j}f(r_{ij}).
\end{align}
Here, $\lambda$ is a dimensionless parameter which we need to calculate. The one-body density matrix can be written in terms of this new trial wave function as
\begin{align}\label{eq:feenberdensitymatrix}
    &n(r_{11'}) =\\\nonumber &\quad\quad n \ee^{-\lambda}\int \dd \bm R_1 |\Psi^{(N-1)}(2,\ldots,N)|^2 \prod_{j=2}^N f(r_{1j})f(r_{1'j}).
\end{align}

The normalization parameter $\lambda$ can be computed from \eqref{eq:feenberjastrow} by comparing a wave function for $N$ and for $N-1$ bosons, see Ref.~\cite{feenberg1974gsi} for details. From this comparison, $\lambda$ can be calculated in several orders of the previously discussed cluster function $h(r)$. Up to second order in the cluster function we get
\begin{align}
    \lambda = D^{[1]}[h] + D^{[2]}[h] + \ldots  = D[h]\;,
\end{align}
where $D^{[1]}[h]$ and $D^{[2]}[h]$ are functionals of $h$ of first and second order, given by,
\begin{align}
    D^{[1]}[h] &= n\int\dd\bm r h(r),\\
    D^{[2]}[h] &= \frac{n^2}{2}\int\dd\bm r_2\dd\bm r_3\;g(r_{23}-1)h(r_{12})h(r_{13})\;.
\end{align}

The expression for the density matrix in \eqref{eq:feenberdensitymatrix} can be expanded in a similar way. However, instead of the cluster function $h(r)$, the radial function $\zeta(r) = f(r)-1$ is used. This function has the similar property that it goes (quickly) to zero for large $r$, and hence we can also perform a cluster expansion. Since the normalization was computed with four copies of $f$, and thus second order in $h$, the density matrix is also computed with four copies of $f$ and thus to fourth order in $\zeta$.

Again, the precise details of this calculation can be found in Ref.~\cite{ristig1979lob}, but up to second order in $f^2$ the result is,
\begin{align}\begin{split}
    n^{[2]} = n \ee^{\lambda^{[2]}}&\exp{\left(2D^{[1]}[\zeta]-Q^{[1]}(r) \right) }\times \\
          &\quad\quad\times \exp{\left(2D^{[2]}[\zeta]-Q^{[2]}(r) \right) }\;,
\end{split}\end{align}
which in general can be written as,
\begin{align}\label{eq:densmatrixQ}
    n(r) = n \exp{\left(2D[\zeta]-D[h]-Q(r)\right)}\;,
\end{align}
with $Q^{[1]}(r)+Q^{[2]}(r)+\ldots=Q(r)$. The function $Q(r)$ contains every term that still depends on $|\bm r_1-\bm r_1'|$, all of which go to zero for $r\rightarrow\infty$. All constant terms turn out to have the same functional form as the normalization terms, and can be expressed in the same functional $D$. If we compare \eqref{eq:densmatrixQ} with \eqref{eq:densmatrix} we notice that it has exactly the same form. Thus when we take the limit $r\rightarrow\infty$ we get for the condensate density,
\begin{align}
    n_0 = n\exp{\left(2D[\zeta]-D[h]\right)}\;.
\end{align}
When we insert the expression for $D$ up to second order in $f^2$, we get
\begin{align}\label{eq:condensfraction}
    &n_0 = n\exp \bigg[ -n\int\dd\bm r\zeta(r)^2  \\\nonumber
    &\quad\quad\quad +n\int\frac{\dd\bm k}{^(2\pi)^3}(S(k) -1)\left(\zeta(k)^2-\frac{1}{2}h(k)^2\right)\bigg]\;,
\end{align}
here we have used the Fourier transform of $\zeta$  and $h$ in order to get rid of double integrals over $r$. This expression for the condensate density only depends on $f$ and $g$ and we are now able to compute this for the minimized results below.

\section{Variational solutions}\label{sec:variational}

With the HNC approximation for the Jastrow wave function, we have an expression for the energy in terms of the two-particle distribution function. The ground-state $g(r)$ minimizes this energy. In the previous section we derived a differential equation for $g(r)$ in \eqref{eq:minimizeg} that solves the minimization equation. However, this is a very nonlinear equation in $g(r)$, since the effective potential $\omega_0(r)$ in \eqref{eq:defomega} depends on $g(r)$ in a complicated way. This makes solving the differential equation very difficult. A variational approach, where we directly vary $g(r)$ to find an energy minimum, turns out to work much better.

\subsection{Potential with resonance}\label{sec:feshbach}

In the unitarity limit, it is expected that the system behaves universally; this behavior does therefore not depend on the exact shape of the interaction potential. This gives us the possibility to choose a simple potential that is numerically convenient, and also contains a `Feshbach' resonance to go to the unitarity limit. The potential we choose is a hard core combined with an attractive $1/r^6$ tail. One of the advantages of this potential is that the two-particle problem can be solved exactly. From the two-particle solutions, the scattering length can be determined, which diverges for certain values of the interaction strength of the attractive tail.

\begin{figure}
    \includegraphics[width=.9\columnwidth]{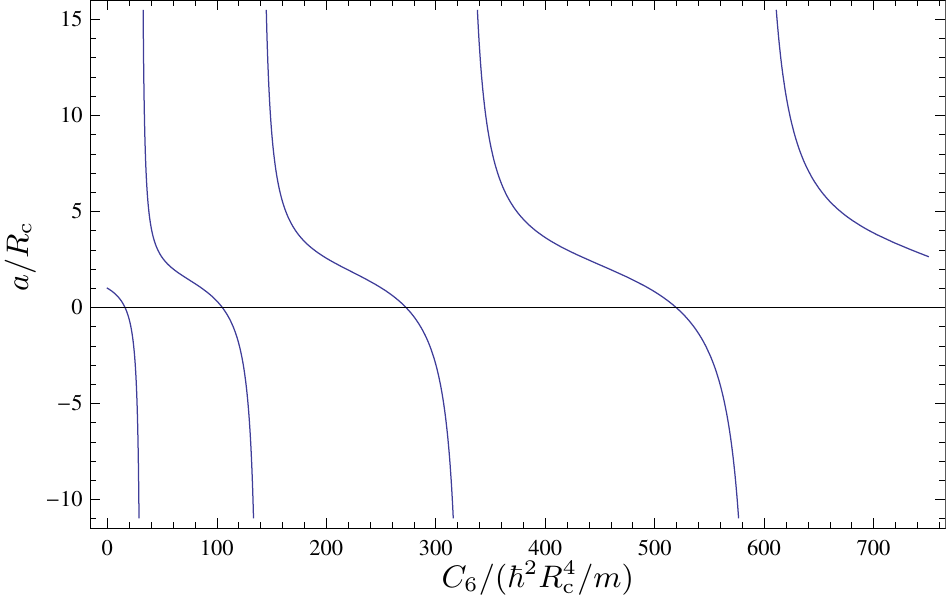}
    \caption{(Color online) The scattering-length $a$ in units of $R_{\rm c}$ as a function of the dimensionless interaction strength $C_6 / (\hbar^2 R_{\rm c}^4/m)$. At certain values of $C_6$ the scattering length diverges and the system is at a resonance.}\label{fig:feshbachresonance}
\end{figure}

The interaction potential has a hard core with radius $R_{\rm c}$, and has the following form,
\begin{align}
    V(r) =
    \begin{cases}
        \infty & r<R_{\rm c} \\
        -\frac{C_6}{r^6} & r \geq R_{\rm c}
    \end{cases} \ .
\end{align}
This potential is spherically symmetric and since we are looking at dilute and ultracold gases, only the \emph{s}-wave part of the interaction is important. In the spherically symmetric case it is convenient to define
\begin{align}
    f_2(r) = \frac{u(r)}{r} \ ,
\end{align}
where $f_2(r)$ is the two-particle wave function. The Schr\"odinger equation for $u(r)$ can be written as,
\begin{align}
   \left( \frac{\hbar^2}{m}\frac{\dd^2}{\dd r^2} + \frac{C_6}{r^6} \right) u(r) = 0 \ .
\end{align}
This differential equation is solved by
\begin{align}
    u(r) = \sqrt{r} \left[c_1 J_{-\frac{1}{4}}\left(\frac{\sqrt{m C_6}}{2 \hbar r^2}\right)+c_2
   J_{\frac{1}{4}}\left(\frac{\sqrt{m C_6}}{2 \hbar r^2}\right)\right],
\end{align}
where $J_{\pm\frac{1}{4}}(r)$ is the Bessel function of the first kind. The hard core is included with the boundary condition $u(R_{\rm c}) = 0$ and the normalization of the wave function demands that $f_2(r\rightarrow\infty) = 1$. These two relations fix the constants $c_1$ and $c_2$ and we obtain
\begin{align}
    &f_2(r) =\\\nonumber
    & \ \   \frac{ \left[J_{-\frac{1}{4}}\left(\frac{\sqrt{m C_6}}{2  \hbar r^2}\right)
   J_{\frac{1}{4}}\left(\frac{\sqrt{m C_6}}{2 \hbar R^2_{\rm c}}\right)-J_{\frac{1}{4}}\left(\frac{\sqrt{m C_6}}{2  \hbar r^2}\right)
   J_{-\frac{1}{4}}\left(\frac{\sqrt{m C_6}}{2 \hbar  R^2_{\rm c}}\right)\right]}{\sqrt{2} \Gamma \left(\frac{3}{4}\right)^{-1} (m C_6/ \hbar^2)^{-1/8}\sqrt{r} J_{\frac{1}{4}}\left(\frac{\sqrt{m C_6}}{2  \hbar R^2_{\rm c}}\right)}.
\end{align}
From the two-particle wave function we can determine the scattering length $a$,
\begin{align}
    a =
    \frac{1}{2}
    \frac{\Gamma\left(\frac{3}{4}\right)}{\Gamma\left(\frac{5}{4}\right)}\frac{J_{-\frac{1}{4}}\left(\frac{\sqrt{m C_6}}{2\hbar R_{\rm c}^2}\right)}{J_{\frac{1}{4}}\left(\frac{\sqrt{m C_6}}{2\hbar R_{\rm c}^2}\right)}\left(\frac{C_6 m}{\hbar^2}\right)^{1/4}\;.
\end{align}
In \figref{fig:feshbachresonance} the scattering length is plotted as a function of the dimensionless interaction strength $C_6 / (\hbar^2 R_{\rm c}^4/m)$. There clearly are resonances at certain values for $C_6$. We have checked that different shapes of the interaction potential give the same results for the HNC calculation as long as the scattering length is the same. This is due to the dilute limit in which the interaction is governed by the scattering length, and therefore the energy is only a function of $a$ and, in that sense,  independent of $C_6$.

\subsection{Varying the radial distribution function}\label{sec:varrdf}

Within the Jastrow ansatz and the HNC approximation, the energy of the system is completely determined by the two-particle distribution function $g(r)$. In the approach we propose here, we start with an ansatz for $g(r)$ that closely enough resembles the expected functional form, but parametrize enough freedom such that we can find, or get very close to, the actual energy minimum.

The ansatz for $g(r)$ can be constructed out of three parts. The first part is the tail of $g$, that is, the power of $r$ with which $g-1$ approaches zero. As we have seen in \eqref{eq:gtail}, $g(r)$ goes to one for large $r$ as $1-P_4r^{-4}$, where in the weak-coupling limit we also know that $P_4=\hbar/2\pi^2nmc$. The second part of $g$ is the short-range regime. Since HNC incorporates the effects of all particles onto each other, which in this dilute situation is a long-range effect, it has little effect on the short-range behavior of the system. It is therefore reasonable to assume that for small $r$, $g$ is proportional to the two-body function $f_2(r)^2$. The proportionality constant between $f^2_2$ and $g$ is related to the \emph{contact} as discussed in \secref{sec:contact}.  The third part of $g$, which is left over, is the intermediate-range regime. This is where the short- and long-range parts are smoothly connected to each other. More important for this regime is the normalization condition of $g$. This normalization follows directly from the definition of $g(r)$ in \eqref{eq:defineg} and can be written as
\begin{align}
    \int\dd\bm r (1 - g(r)) = 1 \ .
\end{align}
To account for all of this, the middle part needs the most variational freedom. The amount of parameters can be extended by adding a cosine oscillation to the middle part. These oscillations can be expected to be important for the strongly interacting regime, where liquid-like shell structures may occur, although we have not observed this yet.

\begin{figure}[t!]
    \includegraphics[width=.9\columnwidth]{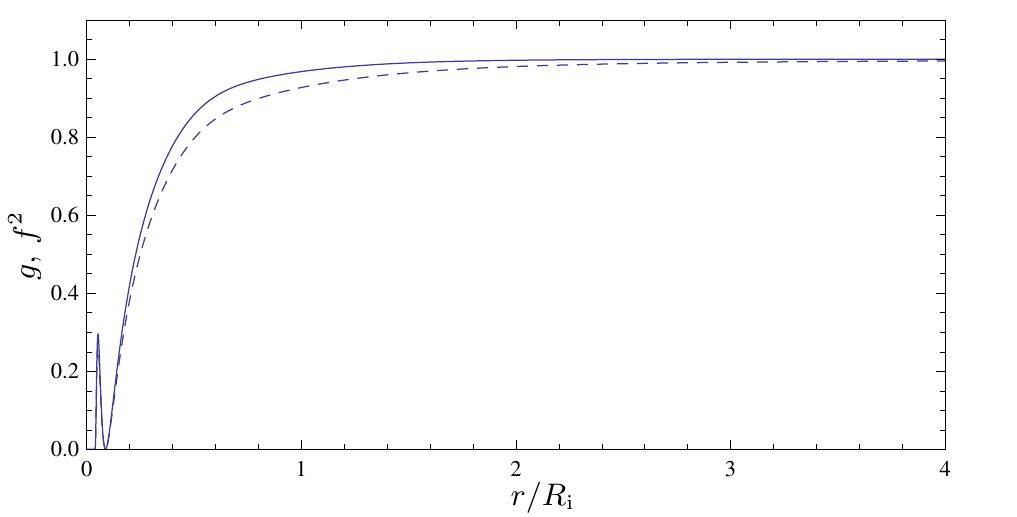}    \includegraphics[width=.9\columnwidth]{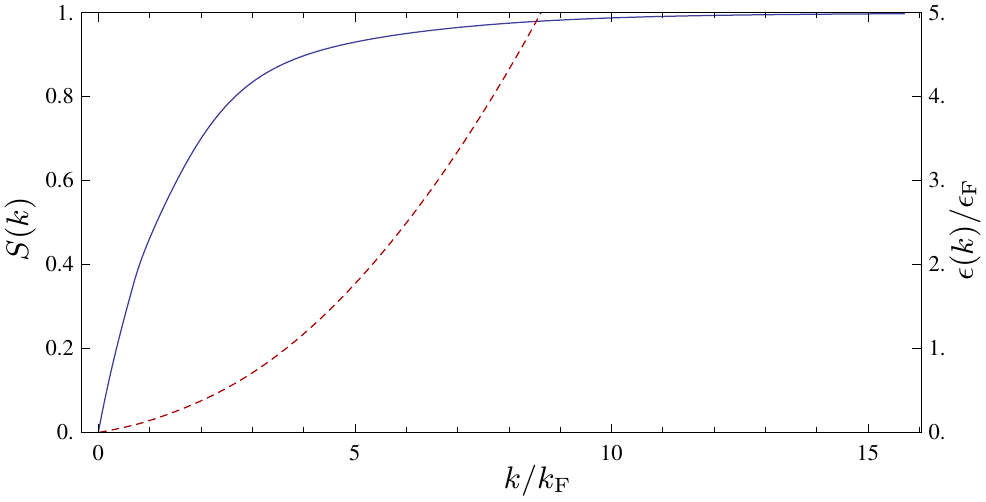}
    \caption{(Color online) (Top) The minimized radial distribution function $g$ (solid line) and $f(r)^2$ (dashed line) as a function of the radius in units of the interparticle distance $R_{\rm i}$. (Bottom) The structure factor $S(k)$ (solid line) and the dispersion relation in units of $\eF$ (dashed line) as a function of the momentum $k/\kF$.}\label{fig:sdispgf2}
\end{figure}

Since $g(r)$ is a distribution function, it is always larger than zero, and we are therefore able to write it as an exponent of another function. Writing it this way has the advantage that it cannot accidentally become negative when varying the parameters. We split up the ansatz for $g(r)$ in a short- ($u_{\rm s}$), a middle- ($u_{\rm m}$), and a long-range ($u_{\rm l}$) part:
\begin{align}\label{eq:ansatzu}
    u_{\rm s} & = \left(2 \log{f_2(r)} + P_8\right) \times \exp{\left(-P_2r^{P_5}\right)} \ ,\\
    u_{\rm m} & = \frac{P_3\cos{\left(P_{10}(r-P_6)\right)}}{P_{3\rm b}+r^{P_1}} \times \left(1-\exp{\left(-P_9 r^{P_7}\right)}\right),\\
    u_{\rm l} & = \frac{P_4}{P_{4\rm b}+r^4}  \times \left(1-\exp{\left(-P_{11} r^{P_{12}}\right)}\right)\ .
\end{align}
The radial distribution function $g(r)$ is then given by,
\begin{align}\label{eq:ansatzg}
    g(r) = \exp{\left(u_{\rm s} + u_{\rm m} + u_{\rm l}\right)} \ .
\end{align}
This parametrization of the radial distribution function was common practice in the field of liquid $^4$He, as for instance in Ref.~\cite{smith1979gsb}.

\subsection{Results}\label{sec:results}

In the previous sections we showed how a Jastrow wave function, together with the HNC approximation, can be used to compute several properties of a Bose gas toward a Feshbach resonance. In this section we will show the first promising results for small and intermediate scattering lengths.

To find the energy minimum we vary the parameters in the distribution function $g$ in \eqref{eq:ansatzg} and use the HNC approximation to compute the energy. For the variation we use a gradient algorithm which converges slowly toward the energy minimum. For small $\kF a$ this goes relatively fast and easy, but with increasing $\kF a$ it becomes increasingly difficult. We therefore increased the scattering length step by step, and used the resulting parameter values of one minimization as a starting point for the next.

\begin{figure}[t!]
    \includegraphics[width=.9\columnwidth]{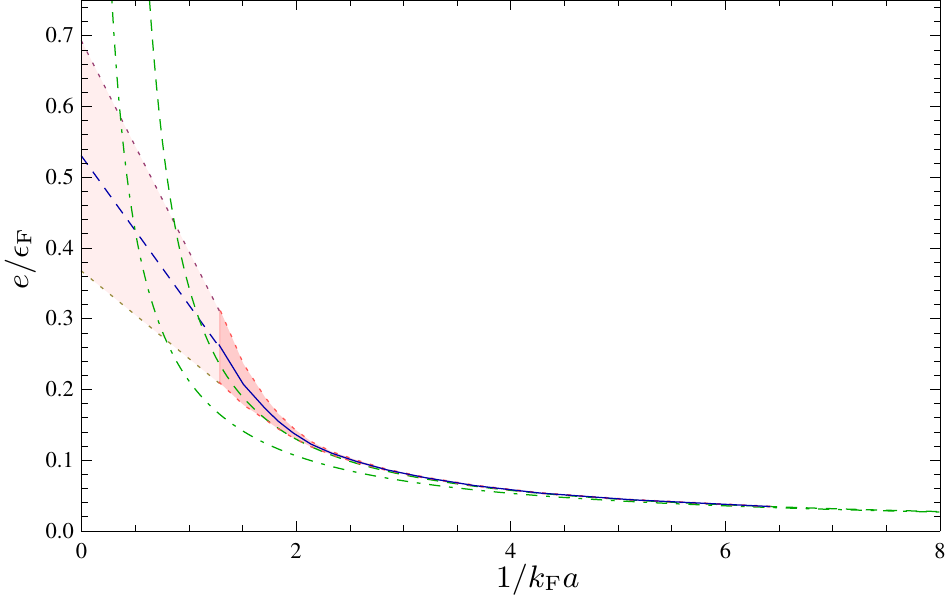}
    \caption{(Color online) The energy as a function of the inverse scattering length $1/\kF a$ calculated with the variational approach of HNC/0 (solid line). The (red-shaded) area depicts the estimated accuracy of the result, which shows the large error for large $\kF a$. The dash-dotted line shows the Gross-Pitaevskii energy while the dashed line also includes the LHY correction in \eqref{eq:LDE}. This shows that the HNC calculation correctly includes this term. }\label{fig:energy}
\end{figure}

In \figref{fig:sdispgf2} we show in the top panel the result of a two-particle distribution function for which the energy is minimized. Notice the wiggle near $r=0$, which shows that we are actually dealing with a meta-stable many-body solution of the used potential, that acts as the ground state in the HNC approximation. We also show $f(r)^2$. In the bottom panel the structure factor is shown (solid line), which is zero for $k=0$, as it is supposed to be. It also starts linearly, and as a result, the dispersion relation (dashed line) also starts linearly for small $k$, but becomes of the usual quadratic shape for larger $k$.

This method works excellently for small and also for intermediate scattering lengths. This can be seen in \figref{fig:energy} where the solid line shows the energy as a function of $\kF a$. For small $\kF a$ ($\kF a\lesssim 0.2$), the energy agrees with the mean-field result in Bogoliubov theory see \eqref{eq:LDE} (dash-dotted line). When we increase $\kF a$ ($0.2\lesssim \kF a \lesssim 0.5$) the energy also includes the LHY correction (dashed line). When $\kF a$ becomes even larger, it becomes increasingly harder to find a reliable energy minimum. To indicate this, we have estimated the accuracy of the energy.

\begin{figure}[t!]
    \includegraphics[width=.9\columnwidth]{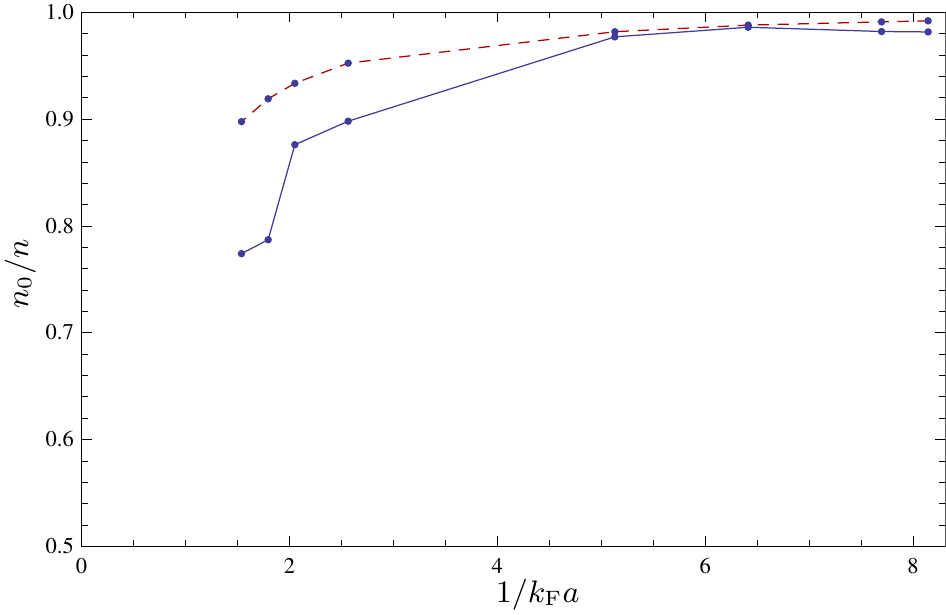}
    \caption{(Color online) The condensate fraction as a function of the inverse scattering length $1/\kF a$ calculated with the variational approach of HNC/0 (solid line). The (red) dashed line shows the condensate fraction for Bogoliubov theory. The condensate fraction for HNC/0 is comparable to the Bogoliubov result for small $\kF a$, but for larger scattering lengths it is significantly smaller.  }\label{fig:depletion}
\end{figure}

Now that we have the two-particle distribution function as a function of the scattering length that minimizes the energy, we can calculate several other physical quantities. One such quantity is the condensate fraction. In \eqref{eq:condensfraction} we showed how this condensate fraction can be calculated given the radial distribution function. In \figref{fig:depletion} the condensate fraction is plotted as a function of the inverse scattering length $1/\kF a$. The blue solid line is the result from HNC/0 and the red dashed line is the result from Bogoliubov theory. The result from HNC/0 is comparable to the Bogoliubov result for small scattering lengths, but for larger values of $\kF a$ the depletion in the HNC/0 case is significantly higher than for Bogoliubov theory.

The contact, which was discussed in \secref{sec:contact} is also an important physical quantity. We showed two ways to extract the contact from the HNC/0 results: one directly from the two-particle distribution function \eqref{eq:contactfromg}; the other as a derivative of the energy \eqref{eq:tancontact}. Since the energy follows the Bogoliubov energy, the contact computed as the derivative of the energy with respect to $1/a$, is roughly the same as the Bogoliubov contact in \eqref{eq:ldecontact}. However, since convergence is not properly reached for some of the large-$\kF a$ points in \figref{fig:energy}, the contact cannot be computed there either. Furthermore, since one expects the energy to be finite at unitarity, the contact should also be finite in that regime.

\begin{figure}[t!]
    \includegraphics[width=.9\columnwidth]{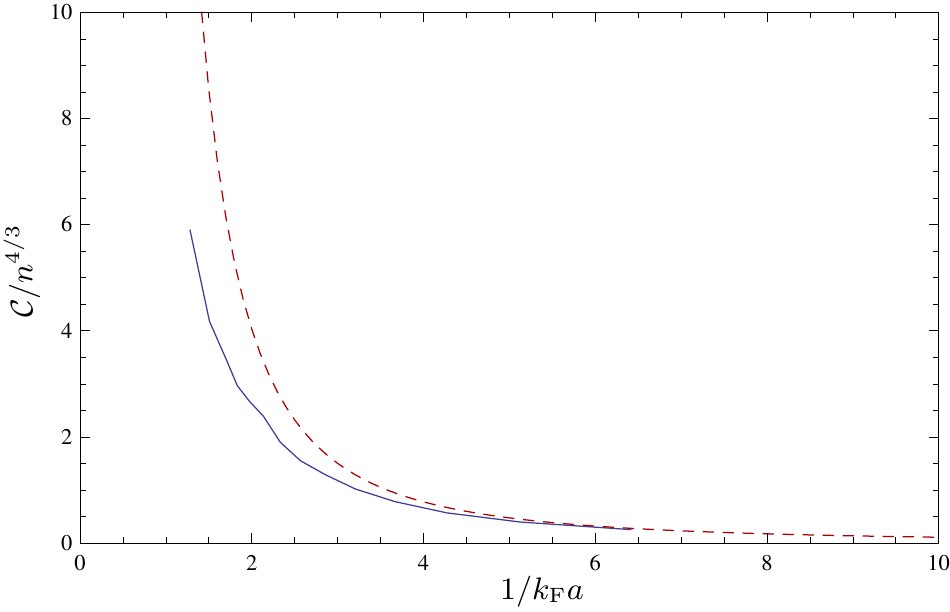}
    \caption{(Color online) The contact as a function of the inverse scattering length $1/\kF a$ calculated with the variational approach of HNC/0 (solid line). The (red) dashed line shows the contact for the Bogoliubov theory in \eqref{eq:ldecontact}.  }\label{fig:contact}
\end{figure}

The second method, where we directly read off the contact from $g$, might indicate already that the contact becomes smaller than the Bogoliubov result, which is shown in \figref{fig:contact}. Also, the slow convergence of the variational process prevents us from computing the contact up to the unitarity limit, but the results for intermediate $\kF a$ show a decrease in $\mathcal{C}$. The fact that the contact is smaller also indicates that the wave function renormalization constant $Z$ is smaller than one. This would indicate that the three-body particle decay rate is suppressed by many-body effects in the unitarity limit.

\section{Conclusion}\label{sec:conclusion}

The unitary regime for bosons is still not completely understood, both experimentally and theoretically. In this paper we believe to have shown that the use of a Jastrow ansatz with the HNC approximation gives promising results that will help with the understanding.

In the first section we put forward a very elegant mean-field theory which describes a Bose gas near a Feshbach resonance. This theory shows the universal nature at a Feshbach resonance and can be used to calculate numerically the chemical potential as a function of the scattering length. Moreover, at unitarity this theory gives an analytic result for the chemical potential.

This mean-field theory is probably, for large interaction strengths, quantitatively not reliable, since it lacks the important contributions of quantum fluctuations. We therefore propose to use a Jastrow ansatz together with the HNC approximation to further investigate the strongly interacting Bose gas. We have shown how to set up such an approach. From the two-particle distribution function $g$, which is computed with the hypernetted-chain relation, several important physical quantities can be derived. Not only the energy, but also the condensate fraction and the contact can be computed directly from $g$.

The system of relations for $f$ and $g$ can be solved using a variational approach. The two-particle distribution function is varied, until the energy is minimized. This gives promising results for small and intermediate values of the scattering length $\kF a$. For larger $\kF a$, the chosen parametrization of $g$ does not converge in a stable manner. It is yet unclear whether this is purely a numerical problem or if there are real physical instabilities involved. Further work is needed to fully understand this issue. However, for the regime where convergence is reached, we were able to derive the energy, the condensate fraction, and the contact.

The ultimate goal would obviously be to find the energy exactly at unitarity. To give a first estimate, we show in \figref{fig:energy} with the dashed blue and red lines an extrapolation of the energy. From the mean-field result in \secref{sec:mf}, we notice that the energy leaves linearly in $1/\kF a$ from unitarity. This also corresponds to a constant contact at unitarity, which is related to the slope of the energy. When we assume this behavior to be correct, we find an energy $e \simeq 0.5 \eF$. At unitarity the chemical potential is related to the energy via $\mu = 5 e / 3$, which results in an estimate of the universal number $\beta\simeq-0.2$. This is higher than the $\beta=-0.54$ found using a mean-field theory; however, since the convergence for the higher $\kF a$ energy could not be reached completely, we expect this $\beta$ to be an upper bound. This is in agreement with the current experiments and calculations \cite{navon2011dtl,lee2010usd,song209gsp,cowell2002cbg}. For the contact we find at unitarity $\mathcal{C}\simeq10.3\, n^{4/3}$, which is remarkably close to the contact for unitary fermions, $\mathcal{C}\simeq11\, n^{4/3}$ \cite{tan2008lmp}. This poses again the interesting question whether the universal behavior of fermions and bosons is identical at unitarity.

\subsection*{Acknowledgements}
We would like to thank Misha Veldhoen, Erik van der Bijl and Randy Hulet for the interesting and fruitful discussions.

\end{document}